\documentstyle[12pt]{article}
\title{\bf Casimir effect in a two dimensional signature changing spacetime}
\author{F. Darabi \thanks{Corresponding author e-mail:f.darabi@azaruniv.edu} and
M. R. Setare \thanks{e-mail:rezakord@ipm.ir}\\
{\small Department of Physics, Azarbaijan University of Tarbiat
Moallem, 53714-161 Tabriz, Iran .}\\
{\small Physics Dept. Inst. for Studies in Theo. Physics and
Mathematics (IPM), 19395-5531, Tehran, Iran .} }
\begin{document}
\maketitle
\vspace{15mm}

\begin{abstract}
We study the Casimir effect for free massless scalar fields
propagating on a two-dimensional cylinder with a metric that
admits a change of signature from Lorentzian to Euclidean. We
obtain a nonzero pressure, on the hypersurfaces of signature
change, which destabilizes the signature changing region and so
alters the energy spectrum of scalar fields. The modified region
and spectrum, themselves, back react on the pressure. Moreover,
the central term of diffeomorphism algebra of corresponding
infinite conserved charges changes correspondingly.
\end{abstract}
\newpage

\section{Introduction}

The Casimir effect is usually regarded as the most well-known
manifestation of vacuum fluctuations in quantum field theory. In
this effect, the presence of reflecting boundaries in the quantum
vacuum alters the zero-point modes of the quantized fields, and
results in the shifts of the vacuum expectation values of
quantities such as energy densities and stresses. These shifts
lead to vacuum forces which act on the reflecting boundaries. The
particular features of these forces depend on the nature of the
quantum field, the type of spacetime manifold and its
dimensionality, the boundary geometries and the specific boundary
conditions imposed on the field. Since the original work by
Casimir in 1948 \cite{Casi48} many theoretical and experimental
works have been done on this problem
\cite{Most97,Plun86,Lamo99,Bord99,Bord01,Kirs01,Bord02,Milt02,setare}
. In general, there are several approaches to calculate the
Casimir energy: mode summation \cite{db}, Green's function method
\cite{Plun86}, heat kernel method \cite{Kirs01}, along with
appropriate regularization schemes such as point separation
\cite{chr,adler}, dimensional regularization \cite{deser}, and
zeta function regularization
\cite{{haw},{blu},{Remeo},{Elizalde},{by}}. Recently, general new
methods have been obtained to compute the renormalized one-loop
quantum energies and energy densities \cite{{gram1},{gram2}}.

On the other hand, signature changing spacetimes have recently
been of particular importance as the specific geometries with
interesting physical effects. The original idea of signature
change was due to Hartle, Hawking and Sakharov \cite{HHS}. This
interesting idea would make it possible to have both Euclidean and
Lorentzian metrics in path integral approach to quantum gravity.
Later, it was shown that the signature change may happen in
classical general relativity, as well \cite{CSC}. There are two
different approaches, continuous and discontinuous, to study the
signature change in classical general relativity \cite{CSC,D}. In
the continuous approach, the signature of metric changes
continuously in passing from Euclidean to Lorentzian region.
Hence, the metric becomes degenerate at the border of these
regions. In the discontinuous approach, however, the metric
becomes nondegenerate everywhere and is discontinuous at the
border of Euclidean and Lorentzian regions.

The issue of propagation of quantum fields on signature-changing
spacetimes has also been of some interest \cite{D}. For example,
Dray {\it et al} have shown that the phenomenon of particle
production may happen for scalar particles propagating in a
spacetime with heterotic signature. They have also obtained a rule
for propagation of massless scalar fields on a two dimensional
spacetime with signature change. Dynamical determination of the
metric signature in spacetime of nontrivial topology is another
interesting issue which has been studied in \cite{eor}.

To the authors knowledge, no attempt has been done to study the
Casimir effect within the geometries with signature change. A
relevant work to the present paper is \cite{far}. In this work, a
model of free massless scalar fields on a two dimensional cylinder
with a signature-changing strip has been studied and shown that
the energy spectrum depends on the strip's width and differs from
the ordinary bosonic spectrum, for low energies. Moreover, It was
shown that the diffeomorphism algebra of the corresponding
infinite conserved charges is different from ``Virasoro'' algebra
and approaches it at higher energies.

In this paper, we study the Casimir effect for the free massless
scalar field propagating on the above two-dimensional
signature-changing cylinder. We will obtain a nonzero pressure on
the hypersurfaces of signature change which leads to instability
in the signature-changing region. Therefore, depending on
situation, Euclidean or Lorentzian region will grow or shrink, and
this will alter the energy spectrum and the diffeomorphism algebra
discussed above.

Meanwhile, a lot of topics related to the Casimir effect have been
explored in the context of string theory \cite{Milt02} \cite{STR}.
The above two-dimensional signature-changing cylinder is
topologically similar to a closed string, with two Euclidean and
Lorentzian parts, propagating in a distributional way in the
2-dimensional target space, and the discontinuous nature of the
model in classifying Euclidean and Lorentzian solutions with
discrete symmetry motivates one to study it in the context of {\it
orbifolds} \cite{far} \cite{orb}. On the other hand, a closed
string with two Euclidean and Lorentzian parts may be of some
importance in the context of $D$-branes and related conformal
field theories. Therefore, taking into account this similarity,
the study of Casimir effect in our model may have nontrivial
impacts on closed strings or $D$-branes.

In general, we believe the idea of Casimir effect in
signature-changing spacetimes is novel and interesting. In the
present paper this effect is inevitably limited to two-dimensional
spacetime, which may be relevant to the study of closed bosonic
strings with Euclidean and Lorentzian parts. But, further study of
Casimir effect in 3+1 dimensional signature changing spacetimes
may have more important physical implications, especially at early
universe \cite{SD}.

\section{Casimir stress tensor in signature changing spacetime}

We consider a free massless scalar field $\phi$ which propagates
on a two-dimensional manifold $M= R\times S^{1}$ ( the circle
$S^{1}$ represents {\it space} and the real line $R$ represents
{\it time} ) with the following metric:
\begin{equation}
ds^{2}=-d\tau^{2}+g(\sigma)d\sigma^{2}, \label{met}
\end{equation}
where $\tau$ is timelike coordinate and $\sigma$ is a periodic
spacelike coordinate with the period $L$\footnote{We assume $L$ is
the circumference of the circle with radius $r$. }, and
$g(\sigma)$ is a periodic function of $\sigma$, which takes +1 for
Lorentzian and -1 for Euclidean regions\footnote{Notice that in
this region the metric will be $g_{\alpha \beta}=\mbox{diag}(-1,
-1)$}
\begin{equation}
g(\sigma)=\left \{ \begin{array}{ll}
-1 &\hspace{5mm} 0<\sigma<\sigma_0+ mod\: L ,   \\
+1 &\hspace{5mm} \sigma_0<\sigma<L+ mod\: L ,
\end{array}\right.
\label{g}
\end{equation}
where $\sigma=0, L$ and $\sigma=\sigma_0$ are the hypersurfaces of
signature change. We assume the scalar field to satisfy specific
junction conditions at these hypersurfaces. In the literature of
signature change there are two kinds of junction conditions:\\ i)
$\phi$ and its derivatives are continuous across
$\sigma=\sigma_0$, ({\it Dray et al}) \cite{D},\\ ii) $\phi$ is
continuous but its derivatives vanish across $\sigma=\sigma_0$,
({\it Hayward}) \cite{CSC}.\\ In this paper, we assume the first
junction conditions as the appropriate boundary conditions at each
region. We assume the continuity of $\phi$ as well as its
derivatives at all times $\tau$ as
\begin{equation}
\left \{ \begin{array}{ll} \phi^E\mid_{\Sigma,
\Sigma'}=\phi^L\mid_{\Sigma, \Sigma'},
\\
\\
\partial_{\sigma} \phi^E\mid_{\Sigma, \Sigma'} = - \: \epsilon \; \partial_{\sigma} \phi^L\mid_{\Sigma, \Sigma'},
\end{array}\right.
\label{70}
\end{equation}
where $\Sigma, \Sigma'$ are the hypersurfaces of signature change,
and $\epsilon=\frac{\epsilon^+}{\epsilon^-}$ takes the values $\pm
1$ according to the orientation of the coordinates $\tau$ and
$\sigma$ in both regions of different signatures. Assuming
$\epsilon^+ =+1$ and $\epsilon^- =+1$ for Euclidean and Lorentzian
regions respectively, the junction conditions (\ref{70}) are
written as \cite{far}
\begin{equation}
\left \{ \begin{array}{ll} \phi^E \mid_0 = \phi^L \mid_{L}
\\
\\
\partial_{\sigma} \phi^E \mid_0 = - \partial_{\sigma}\phi^L
\mid_{L},
\end{array}\right.\hspace{10mm}
\left \{ \begin{array}{ll}  \phi^E \mid_{\sigma_0} =\phi^L
\mid_{\sigma_0}
\\
\\
\partial_{\sigma} \phi^E \mid_{\sigma_0} = - \partial_{\sigma} \phi^L
\mid_{\sigma_0}.
\end{array}\right.
\label{10}
\end{equation}
By solving the wave equations
\begin{equation}
\begin{array}{ll}
(\:{\partial_\tau}^2 + {\partial_{\sigma}}^2\:)\:\phi^E(\sigma,\tau)=0,\\
\\
(\:{\partial_\tau}^2 -
{\partial_{\sigma}}^2\:)\:\phi^L(\sigma,\tau) =0,
\end{array}
\end{equation}
in both Euclidean and Lorentzian regions and imposing the junction
conditions (\ref{10}), we obtain nontrivial solutions for
$\phi_\omega$, provided the continuous spectrum $\omega$ satisfies
the following ``quantization condition''\footnote{The spectrum
$\omega$ in this model is obtained by solving the quantization
condition which leads to real and $\sigma_0$-dependent values. It
differs from ordinary spectrum (with pure Lorentzian signature) at
low energies and coincides with the integer roots of
$\cos{\omega(L-\sigma_0)}$, at high energies. Therefore, ``sum
over energies'' approaches ``sum over integers'' at higher
energies \cite{far}.}\cite{far}
\begin{equation}
\cosh {\omega \sigma_0}\: \cos {\omega (\sigma_0 - L)} =1.
\label{11}
\end{equation}
It is shown in \cite{far}, that due to the same {\it time
evolution} of the functions $\Phi_\omega^E$ and $\Phi_\omega^L$
one can construct a set of real {\it distributional} orthogonal
and complete solutions on the arbitrary $\tau=const$ hypersurface
as
\begin{equation}
\Phi_\omega(\sigma, \tau)=\Theta^+\:\Phi_\omega^E(\sigma, \tau) +
\Theta^-\:\Phi_\omega^L(\sigma, \tau), \label{23}
\end{equation}
where $\Phi_\omega^E =(\phi^E_\omega + \phi^E_{-\omega}$) ,
$\Phi_\omega^L =(\phi^L_\omega + \phi^L_{-\omega}$), and $\Theta
^{+},\Theta ^{-}$ are Heaviside distributions with support in
Euclidean and Lorentzian regions, respectively\footnote{Heaviside
distributions have the property $ d\Theta ^{\pm }=\pm \delta, $
where $\delta $ is the hypersurface Dirac distribution with
support on the hypersurfaces of signature change.}. The solutions
$\Phi_\omega$ are then expanded as normal mode expansions
\cite{far}.

One can also obtain the following expressions for the components
of energy-momentum tensors associated with the scalar field
$\Phi(\sigma,\tau)$ in both Euclidean and Lorentzian regions
\cite{far}
\begin{equation}
\begin{array}{ll}
T_{00}^E =  [(\partial_\tau\Phi^E)^2 - (\partial_\sigma\Phi^E)^2],
&\hspace{20mm} T_{01}^E
=2\:\partial_\tau\Phi^E\:\partial_\sigma\Phi^E,
\\
\\
T_{00}^L =  [(\partial_\tau\Phi^L)^2 + (\partial_\sigma\Phi^L)^2],
&\hspace{20mm} T_{01}^L
=2\:\partial_\tau\Phi^L\:\partial_\sigma\Phi^L.
\end{array}
\end{equation}
By introducing new coordinates $\sigma_{+}^{E}$, $\sigma_{-}^{E}$
in the Euclidean region, and $\sigma_{+}^{L}$, $\sigma_{-}^{L}$ in
the Lorentzian one as
\begin{equation}
\begin{array}{ll}
\sigma_{+}^{E}=\tau+i\sigma,
&\hspace{20mm}\sigma_{+}^{L}=\tau+\sigma,
\\
\\
\sigma_{-}^{E}=\tau-i\sigma,
&\hspace{20mm}\sigma_{-}^{L}=\tau-\sigma,
\end{array}
\end{equation}
we obtain
\begin{equation}
\begin{array}{ll}
T_{++}^E = (T_{00}^E - iT_{01}^E)/2, &\hspace{20mm} T_{++}^L = (T_{00}^L + T_{01}^L)/2, \\
\\
T_{--}^E = (T_{00}^E + iT_{01}^E)/2, &\hspace{20mm} T_{--}^L = (T_{00}^L - T_{01}^L)/2, \\
\\
T_{+-}^E = T_{-+}^E = 0, &\hspace{20mm} T_{+-}^L = T_{-+}^L = 0.
\end{array}
\label{T}
\end{equation}
Then by substituting the normal mode expansions of the solutions
$\Phi_\omega$ we obtain \cite{far}
\begin{equation}
\begin{array}{l}
T_{++}^E = \frac{1}{2}[(\partial_\tau\Phi - i \partial_\sigma \Phi)]^2\\
 = 2\:\sum_{\omega\omega'} \tilde{f}_\omega^E(+)\:\tilde{f}_{\omega'}^E(+) \:\tilde{\alpha}_\omega\:\tilde{\alpha}_{\omega'}^\dag,\\
\\
T_{--}^E = \frac{1}{2}[(\partial_\tau\Phi + i \partial_\sigma \Phi)]^2 \\
= 2\:\sum_{\omega \omega'} \tilde{f}_{-\omega}^E(-)\:\tilde{f}_{-\omega'}^E(-) \:\tilde{\alpha}_\omega\:\tilde{\alpha}_{\omega'}^\dag,\\
\\
T_{++}^L = \frac{1}{2}[(\partial_\tau\Phi +  \partial_\sigma \Phi)]^2 \\
= 2\:\sum_{\omega\omega'} \tilde{f}_\omega^L(+)\:\tilde{f}_{\omega'}^L(+) \:\tilde{\alpha}_\omega\:\tilde{\alpha}_{\omega'}^\dag,\\
\\
T_{--}^L = \frac{1}{2}[(\partial_\tau\Phi -  \partial_\sigma
\Phi)]^2\\ = 2\:\sum_{\omega\omega'}
\tilde{f}_{-\omega}^L(-)\:\tilde{f}_{-\omega'}^L(-)
\:\tilde{\alpha}_\omega\:\tilde{\alpha}_{\omega'}^\dag, \label{nm}
\end{array}
\end{equation}
where $\tilde{\alpha}^\dag$ and $\tilde{\alpha}$ are the creation
and annihilation operators, respectively, and \footnote{The scalar
product $<\Phi_\omega,\Phi_\omega>$ has been defined in
\cite{far}}
\begin{equation}
\begin{array}{ll}
\tilde{f}_\omega^E(\pm) = [(a/b)_\omega +1] \exp(- i \omega{\sigma^E_\pm})/\sqrt{4\pi<\Phi_\omega,\Phi_\omega>},\\
\\
\tilde{f}_\omega^L(\pm) = [(c/b)_\omega + (1/b)_{-\omega}]\exp(-
i\omega{\sigma^L_\pm})/\sqrt{4\pi<\Phi_\omega,\Phi_\omega>},
\end{array}
\label{f}
\end{equation}
with $(a/b)_\omega$,$(c/b)_\omega$,  $(1/b)_\omega$ given by
\cite{far}
\begin{equation}
\begin{array}{ll}
(a/b)_\omega =  \frac{sin\omega(\sigma_0-L)}{cosh\omega\sigma_0 + sinh\omega\sigma_0 - cos\omega(\sigma_0-L)}, \\
\\
(c/b)_\omega =\frac{(1+i)exp(2\pi i\omega)(sinh\omega\sigma_0 + cosh\omega\sigma_0 - exp(i\omega(\sigma_0-L)))}{2(cosh\omega\sigma_0 + sinh\omega\sigma_0 - cos\omega(\sigma_0-L))},\\
\\
(1/b)_\omega=\frac{(1-i)exp(-2\pi i\omega)(sinh\omega\sigma_0 + cosh\omega\sigma_0 - exp(-i\omega(\sigma_0-L)))}{2(cosh\omega\sigma_0 + sinh\omega\sigma_0 - cos\omega(\sigma_0-L))}.\\
\end{array}
\end{equation}
Notice that in obtaining these results we have imposed the
quantization condition (\ref{11}). \\Now, substituting the normal
mode expansions (\ref{nm}) into Eqs.(\ref{T}) leads to
\begin{equation}
\begin{array}{ll}
T_{00}^E=T_{++}^E + T_{--}^E =2\:\sum_{\omega\omega'} (
\tilde{f}_\omega^E(+)\:\tilde{f}_{\omega'}^E(+) +
\tilde{f}_{-\omega}^E(-)\:\tilde{f}_{-\omega'}^E(-) )
\:\tilde{\alpha}_\omega\:\tilde{\alpha}_{\omega'}^\dag, \\
\\
T_{01}^E=i(T_{++}^E - T_{--}^E) =2i \:\sum_{\omega\omega'} (
\tilde{f}_\omega^E(+)\:\tilde{f}_{\omega'}^E(+) -
\tilde{f}_{-\omega}^E(-)\:\tilde{f}_{-\omega'}^E(-) )
\:\tilde{\alpha}_\omega\:\tilde{\alpha}_{\omega'}^\dag,\\
\\
T_{00}^L=T_{++}^L + T_{--}^L=2\:\sum_{\omega\omega'} (
\tilde{f}_\omega^L(+)\:\tilde{f}_{\omega'}^L(+) +
\tilde{f}_{-\omega}^L(-)\:\tilde{f}_{-\omega'}^L(-) )
\:\tilde{\alpha}_\omega\:\tilde{\alpha}_{\omega'}^\dag,\\
\\
 T_{01}^L=T_{++}^L - T_{--}^L=2\:\sum_{\omega\omega'} (
\tilde{f}_\omega^L(+)\:\tilde{f}_{\omega'}^L(+) -
\tilde{f}_{-\omega}^L(-)\:\tilde{f}_{-\omega'}^L(-) )
\:\tilde{\alpha}_\omega\:\tilde{\alpha}_{\omega'}^\dag.
\end{array}
\end{equation}

\section{VEV of Casimir stress tensor }

The poisson bracket structure for $\Phi$ and its conjugate
momentum $\Pi$ is given by equal time relations
\begin{equation}
\begin{array}{ll}
\{\Phi(\sigma),\Pi(\sigma^\prime)\} =  \delta(\sigma-\sigma^\prime),\\
\\
\{\Phi(\sigma),\Phi(\sigma')\} = \{\Pi(\sigma),\Pi(\sigma')\} = 0.
\end{array}
\end{equation}
By substituting the normal mode expansions of $\Phi$ and $\Pi$
together with the expansion of $\delta(\sigma-\sigma^\prime)$ we
obtain \cite{far}
\begin{equation}
[\tilde{\alpha}_\omega,\tilde{\alpha}_{\omega'}^\dag] = \omega
\delta_{\omega+\omega',0}.
\end{equation}
If we assume $|0_L>$ to be the vacuum state for the signature
changing cylinder $R\times S^{1}$ with circumference $L$, then
$$
\tilde{\alpha}_\omega|0_L>=0,
$$
$$
<0_L|\tilde{\alpha}_{\omega}^\dag=0,
$$
\begin{equation}
 <0_L|\tilde{\alpha}_\omega
\tilde{\alpha}^\dag_{\omega'}|0_L>=\omega
\delta_{\omega+\omega',0}
\end{equation}
and we obtain the following vacuum expectation values for the
components of energy momentum tensors
 \begin{equation}
\begin{array}{ll}
 <0_L|T_{00}^E|0_L>=2\:\sum_{\omega\omega'} (
\tilde{f}_\omega^E(+)\:\tilde{f}_{\omega'}^E(+) +
\tilde{f}_{-\omega}^E(-)\:\tilde{f}_{-\omega'}^E(-) ) \omega
\delta_{\omega+\omega',0} ,
\\
\\
<0_L|T_{01}^E|0_L>=2i \:\sum_{\omega\omega'} (
\tilde{f}_\omega^E(+)\:\tilde{f}_{\omega'}^E(+) -
\tilde{f}_{-\omega}^E(-)\:\tilde{f}_{-\omega'}^E(-) ) \omega
\delta_{\omega+\omega',0} ,
\\
\\
<0_L|T_{00}^L|0_L>=2\:\sum_{\omega\omega'} (
\tilde{f}_\omega^L(+)\:\tilde{f}_{\omega'}^L(+) +
\tilde{f}_{-\omega}^L(-)\:\tilde{f}_{-\omega'}^L(-) ) \omega
\delta_{\omega+\omega',0} ,
\\
\\
<0_L|T_{01}^L|0_L>=2\:\sum_{\omega\omega'} (
\tilde{f}_\omega^L(+)\:\tilde{f}_{\omega'}^L(+) -
\tilde{f}_{-\omega}^L(-)\:\tilde{f}_{-\omega'}^L(-) ) \omega
\delta_{\omega+\omega',0} ,
\\
\\
<0_L|T_{11}^E|0_L>=-<0_L|T_{00}^E|0_L>,
\\
\\
<0_L|T_{11}^L|0_L>=<0_L|T_{00}^L|0_L>.
 \end{array}
\end{equation}
Now, by the following normalization
\begin{equation}
4\pi<\Phi_{\omega}, \Phi_{\omega}>=\frac{1}{2}(L-\sigma_0),
\end{equation}
and inserting it into equations (\ref{f}) and then applying the
delta function $\delta_{\omega+\omega',0}$ we obtain
\begin{equation}
\begin{array}{ll}
<0_L|T_{00}^E|0_L>=\frac{2}{L-\sigma_0} \:\sum_{\omega=0}^{\infty}
[ (\frac{a}{b})_{\omega}+1][(\frac{a}{b})_{-\omega}+1]\omega,
\\
\\
<0_L|T_{01}^E|0_L>=<0_L|T_{10}^E|0_L>=0,
\\
\\
<0_L|T_{11}^E|0_L>=-<0_L|T_{00}^E|0_L>,
\\
\\
<0_L|T_{00}^L|0_L>=\frac{2}{L-\sigma_0} \:\sum_{\omega=0}^{\infty}
[
(\frac{c}{b})_{\omega}+(\frac{1}{b})_{-\omega}][(\frac{c}{b})_{\omega}+(\frac{1}{b})_{-\omega}]^{*}\omega,
\\
\\
<0_L|T_{01}^L|0_L>=<0_L|T_{10}^L|0_L>=0,
\\
\\
<0_L|T_{11}^L|0_L>=<0_L|T_{00}^L|0_L>. \label{T11}
\end{array}
\end{equation}
By using the Heaviside distribution $\Theta ^{+},\Theta ^{-}$ we
may write
\begin{equation}
\begin{array}{ll}
<0_L|T_{00}|0_L>=\Theta^+<0_L|T_{00}^E|0_L>+\Theta^-<0_L|T_{00}^L|0_L>,
\\
\\
<0_L|T_{11}|0_L>=\Theta^+<0_L|T_{11}^E|0_L>+\Theta^-<0_L|T_{11}^L|0_L>.
\end{array}
\end{equation}
The normal ordered expressions are then as follows
\begin{equation}
\begin{array}{ll}
<0_L|:T_{00}:|0_L>=\Theta^+<0_L|:T_{00}^E:|0_L>+\Theta^-<0_L|:T_{00}^L:|0_L>,
\\
\\
<0_L|:T_{11}:|0_L>=\Theta^+<0_L|:T_{11}^E:|0_L>+\Theta^-<0_L|:T_{11}^L:|0_L>,
\end{array}
\label{*}
\end{equation}
such that
\begin{equation}
\begin{array}{ll}
<0_L|:T_{00}^E:|0_L>=<0_L|T_{00}^E|0_L>-\lim_{L^\prime \rightarrow
\infty}<0_{L^\prime}|T_{00}^E|0_{L^\prime}>,
\\
\\
<0_L|:T_{00}^L:|0_L>=<0_L|T_{00}^L|0_L>-\lim_{L^\prime \rightarrow
\infty}<0_{L^\prime}|T_{00}^L|0_{L^\prime}>,
\\
\\
<0_L|:T_{11}^E:|0_L>=<0_L|T_{11}^E|0_L>-\lim_{L^\prime \rightarrow
\infty}<0_{L^\prime}|T_{11}^E|0_{L^\prime}>,
\\
\\
<0_L|:T_{11}^L:|0_L>=<0_L|T_{11}^L|0_L>-\lim_{L^\prime \rightarrow
\infty}<0_{L^\prime}|T_{11}^L|0_{L^\prime}>,
\end{array}\label{**}
\end{equation}
where the second terms are introduced to remove the ultraviolet
divergences in the first terms \cite{db}. Since we have Euclidean
and Lorentzian quantities in the second terms, then the state
$|0_{L^\prime}>$ should have the property:
$|0_{L^\prime}>\rightarrow|0>$ as $L^\prime\rightarrow \infty$,
such that $|0>$ is the vacuum state of an $R\times R$ signature
changing spacetime. The Casimir energy is measured with respect to
this state.

\section{Regularization of VEV of Casimir stress tensor }

Because both terms on the r.h.s of above equations are
individually divergent they have to be subtracted by careful
analysis. By introducing the cut-off to the sums in $00$
components we obtain
\begin{equation}
\begin{array}{ll}
<0_L|T_{00}^E|0_L>_{cut-off}=\frac{2}{L-\sigma_0}
\:\sum_{\omega=0}^{\infty} [
(\frac{a}{b})_{\omega}+1][(\frac{a}{b})_{-\omega}+1]\omega
e^{-\alpha \omega},
\\
\\
<0_L|T_{00}^L|0_L>_{cut-off}= \frac{2}{L-\sigma_0}
\:\sum_{\omega=0}^{\infty} [
(\frac{c}{b})_{\omega}+(\frac{1}{b})_{-\omega}][(\frac{c}{b})_{\omega}+(\frac{1}{b})_{-\omega}]^{*}\omega
e^{-\alpha \omega}.
\end{array}
\end{equation}
The $11$ components will be easily obtained in terms of $00$
components through Eqs.(\ref{T11}). We now break down each sum
into two separate sums
\begin{equation}
\begin{array}{ll}
<0_L|T_{00}^E|0_L>_{cut-off}=\frac{2}{L-\sigma_0}
\:\sum_{\omega<N} [
(\frac{a}{b})_{\omega}+1][(\frac{a}{b})_{-\omega}+1]\omega
e^{-\alpha \omega}
\\
\\
+\frac{2}{L-\sigma_0} \:\sum_{\omega=N}^{\infty} [
(\frac{a}{b})_{\omega}+1][(\frac{a}{b})_{-\omega}+1]\omega
e^{-\alpha \omega},
\\
\\
<0_L|T_{00}^L|0_L>_{cut-off}= \frac{2}{L-\sigma_0}
\:\sum_{\omega<N} [
(\frac{c}{b})_{\omega}+(\frac{1}{b})_{-\omega}][(\frac{c}{b})_{\omega}+(\frac{1}{b})_{-\omega}]^{*}\omega
e^{-\alpha \omega}
\\
\\
+ \frac{2}{L-\sigma_0} \:\sum_{\omega=N}^{\infty} [
(\frac{c}{b})_{\omega}+(\frac{1}{b})_{-\omega}][(\frac{c}{b})_{\omega}+(\frac{1}{b})_{-\omega}]^{*}\omega
e^{-\alpha \omega},
\end{array}
\end{equation}
where $N$ is the root of quantization condition so that for any
$\omega\geq N$ the spectrum almost coincides with integer values.
As is shown in Fig.2 in \cite{far}, it is easily seen from
quantization condition that whatever can $\sigma_0$ be, the real
spectrum of $\omega$ approaches the integer one, generally at
higher values of $N$. Hence, we assume $N$ to be sufficiently
large so that the first sums over $\omega< N$ be (finite) {\it
sums over real values} and the second sums over $\omega\geq N$ be
almost (infinite) {\it sums over integer values}. Thus, we may
discard the cut-off $e^{-\alpha \omega}$ from the finite sums and
keep them just for infinite sums.

Now, we consider the second sum in $<0_L|T_{00}^E|0_L>_{cut-off}$.
Since each of the terms $[ (\frac{a}{b})_{\omega}+1]$ and
$[(\frac{a}{b})_{-\omega}+1]$ approaches 1 for large $\omega$ (
because $(\frac{a}{b})_{\pm \omega}$ almost vanish for large
$\omega$ ) then this sum goes like \begin{equation}
\frac{2}{L-\sigma_0} \:\sum_{\omega=N}^{\infty} \omega e^{-\alpha
\omega}. \label{sumE}
\end{equation}
In the same way for $<0_L|T_{00}^L|0_L>_{cut-off}$, it is easily
shown that each of the terms $[
(\frac{c}{b})_{\omega}+(\frac{1}{b})_{-\omega}]$ and
$[(\frac{c}{b})_{\omega}+(\frac{1}{b})_{-\omega}]^*$ approaches 1
for large integer-like values $\omega\geq N$, and the second sum
goes like
\begin{equation}
\frac{2}{L-\sigma_0} \:\sum_{\omega=N}^{\infty} \omega e^{-\alpha
\omega}, \label{sumL}
\end{equation}
as well. Therefore, we calculate this sum for both regions. We
know that $\omega\geq N$ denotes for integers, hence we redefine
$\omega=N$ to (integer) $\Omega=0$. To this end, we note that
$\omega\geq N$ indicates, by definition, the integer roots of
$\cos\omega(L-\sigma_0)=0$ in the quantization condition, from
which we obtain $\omega \geq \frac{n+1/2}{L-\sigma_0}\pi$ and
$N=\frac{n+1/2}{L-\sigma_0}\pi$. This is equal to
$n=\frac{N(L-\sigma_0)}{\pi}-\frac{1}{2}$ with integer $n$.
Therefore, we may define
$\Omega=n-\frac{N(L-\sigma_0)}{\pi}+\frac{1}{2}$, with $\Omega=0,
1, 2, ...\:$ . We also obtain $\omega$ in terms of $\Omega$ as
$\omega=N+\frac{\pi \Omega}{L-\sigma_0}$.

Therefore, the sum (\ref{sumE}) or (\ref{sumL}) in Euclidean and
Lorentzian regions is written as
$$
\frac{2}{L-\sigma_0} \:\sum_{\Omega=0}^{\infty} (N+\frac{\pi
\Omega}{L-\sigma_0})e^{-\alpha(N+\frac{\pi \Omega}{L-\sigma_0})},
$$
which, after some calculations, leads to\footnote{We have used
$\sum_{n=0}^{\infty}ne^{-2\pi\alpha
n/L}=e^{2\pi\alpha/L}(e^{2\pi\alpha/L}-1)^{-2}$ \cite{db}.}
\begin{equation}
\frac{2}{L-\sigma_0} \:\sum_{\Omega=0}^{\infty} (N+\frac{\pi
\Omega}{L-\sigma_0})e^{-\alpha(N+\frac{\pi \Omega}{L-\sigma_0})}=
\frac{2}{L-\sigma_0}e^{-\alpha N}\frac{(N+\frac{\pi
\Omega}{L-\sigma_0})e^{-\alpha(N+\frac{\pi
\Omega}{L-\sigma_0})}-N}{(e^{-\alpha(N+\frac{\pi
\Omega}{L-\sigma_0})}-1)^2}.
\end{equation}
First, we focus on the Euclidean calculations. By using
Eqs.(\ref{*}), (\ref{**}) for the Euclidean region we find
$$
\Theta^+<0_L|:T_{00}^E:|0_L>=\Theta^+\{\frac{2}{L-\sigma_0} [
\sum_{\omega=0}^{N-1}
[(\frac{a}{b})_{\omega}+1][(\frac{a}{b})_{-\omega}+1]\omega+
e^{-\alpha N}\frac{(N+\frac{\pi
\Omega}{L-\sigma_0})e^{-\alpha(N+\frac{\pi
\Omega}{L-\sigma_0})}-N}{(e^{-\alpha(N+\frac{\pi
\Omega}{L-\sigma_0})}-1)^2} ]
$$
\begin{equation}
- \frac{2}{L-\sigma_0} [\sum_{\omega=0}^{N-1} \omega + \lim_{L
\rightarrow \infty} e^{-\alpha N}\frac{(N+\frac{\pi
\Omega}{L-\sigma_0})e^{-\alpha(N+\frac{\pi
\Omega}{L-\sigma_0})}-N}{(e^{-\alpha(N+\frac{\pi
\Omega}{L-\sigma_0})}-1)^2}  ] \}, \label{Theta}
\end{equation}
or
$$
\Theta^+\{\frac{2}{L-\sigma_0} [\sum_{\omega=0}^{N-1} [[
(\frac{a}{b})_{\omega}+1][(\frac{a}{b})_{-\omega}+1]-1 ] \omega ]
$$
\begin{equation}
+ \frac{2}{L-\sigma_0} [e^{-\alpha N}\frac{(N+\frac{\pi
\Omega}{L-\sigma_0})e^{-\alpha(N+\frac{\pi
\Omega}{L-\sigma_0})}-N}{(e^{-\alpha(N+\frac{\pi
\Omega}{L-\sigma_0})}-1)^2} - \lim_{L \rightarrow \infty}
e^{-\alpha N}\frac{(N+\frac{\pi
\Omega}{L-\sigma_0})e^{-\alpha(N+\frac{\pi
\Omega}{L-\sigma_0})}-N}{(e^{-\alpha(N+\frac{\pi
\Omega}{L-\sigma_0})}-1)^2}  ] \}, \label{Theta+}
\end{equation}
where the finite sum has appeared without cut-off. In obtaining
$\sum_{\omega=0}^{N-1} \omega$ in the second line of
(\ref{Theta}), we have used $\lim_{L \rightarrow
\infty}(\frac{a}{b})_{\pm\omega}\equiv \lim_{\sigma_0 \rightarrow
\infty}(\frac{a}{b})_{\pm\omega}=0$ in the finite sum. We now
expand the $e^{-\alpha N}\frac{(N+\frac{\pi
\Omega}{L-\sigma_0})e^{-\alpha(N+\frac{\pi
\Omega}{L-\sigma_0})}-N}{(e^{-\alpha(N+\frac{\pi
\Omega}{L-\sigma_0})}-1)^2}$ terms in the second bracket of
(\ref{Theta+}) about $\alpha=0$. After some calculation we obtain
$$
\frac{2}{L-\sigma_0} [e^{-\alpha N}\frac{(N+\frac{\pi
\Omega}{L-\sigma_0})e^{-\alpha(N+\frac{\pi
\Omega}{L-\sigma_0})}-N}{(e^{-\alpha(N+\frac{\pi
\Omega}{L-\sigma_0})}-1)^2}]
$$
$$
= \frac{2}{\alpha^2 \pi}-(\frac{\alpha^2 N^2}{2}-\alpha N +1 )[
\frac{\pi}{6(L-\sigma_0)^2}+\frac{13}{6(L-\sigma_0)}N ]
-\frac{N^2}{\pi}(1-\alpha N ).
$$
Substituting this result into the above bracket and taking $\alpha
\rightarrow 0$ leads to
$$
\frac{2}{L-\sigma_0} [e^{-\alpha N}\frac{(N+\frac{\pi
\Omega}{L-\sigma_0})e^{-\alpha(N+\frac{\pi
\Omega}{L-\sigma_0})}-N}{(e^{-\alpha(N+\frac{\pi
\Omega}{L-\sigma_0})}-1)^2} - \lim_{L \rightarrow \infty}
e^{-\alpha N}\frac{(N+\frac{\pi
\Omega}{L-\sigma_0})e^{-\alpha(N+\frac{\pi
\Omega}{L-\sigma_0})}-N}{(e^{-\alpha(N+\frac{\pi
\Omega}{L-\sigma_0})}-1)^2}  ]
$$
$$
=-\frac{\pi}{6(L-\sigma_0)^2}-\frac{13}{6(L-\sigma_0)}N.
$$
Finally, we have the following expression for the Euclidean region
\begin{equation}
<0_L|:T_{00}^E:|0_L>=\{\frac{2}{L-\sigma_0} \sum_{\omega=0}^{N-1}
[ [ (\frac{a}{b})_{\omega}+1][(\frac{a}{b})_{-\omega}+1]-1 ]
\omega -\frac{\pi}{6(L-\sigma_0)^2}-\frac{13}{6(L-\sigma_0)}N \}.
\label{Euc}
\end{equation}
In the same way, the calculations for the Lorentzian region lead
to
\begin{equation}
<0_L|:T_{00}^L:|0_L>= \{\frac{2}{L-\sigma_0} \sum_{\omega=0}^{N-1}
[[(\frac{c}{b})_{\omega}+(\frac{1}{b})_{-\omega}][(\frac{c}{b})_{\omega}+(\frac{1}{b})_{-\omega}]^*-1
] \omega -\frac{\pi}{6(L-\sigma_0)^2}-\frac{13}{6(L-\sigma_0)}N
\}. \label{Lor}
\end{equation}
Note that, since $N$ is not uniquely determined, then the
expectation values (\ref{Euc}), (\ref{Lor}) are evaluated
approximately. Therefore, for a given $\sigma_0$, replacing $N$ by
$N+1$ or $N-1$ leads to better or worse approximation,
respectively. This is because, as we go to higher values of $N$
the real spectrum coincides with integer one with better
approximation.

\section{Finite energy density and pressure in Euclidean and Lorentzian regions }

By using Eqs.(\ref{T11}) for the $11$ components we have
\begin{equation}
\begin{array}{ll}
<0_L|:T_{11}^E:|0_L>=-<0_L|:T_{00}^E:|0_L>,
\\
\\
<0_L|:T_{11}^L:|0_L>=<0_L|:T_{00}^L:|0_L>.
\end{array}
\end{equation}
Therefore, the state $|0_L>$ contains the finite energy density
and pressure in the Euclidean and Lorentzian regions as follows:
\begin{equation}
\left\{ \begin{array}{ll} \rho^E=<0_L|:T_{00}^E:|0_L>,
\\
\rho^L=<0_L|:T_{00}^L:|0_L>,
\end{array}\right.
\label{ppp}
\end{equation}
\begin{equation}
\left\{\begin{array}{ll} p^E=<0_L|:T_{11}^E:|0_L>=-\rho^E,
\\
p^L=<0_L|:T_{11}^L:|0_L>=\rho^L.
\end{array}\right.
\end{equation}
We then find that the total pressure acting on the signature
changing hypersurfaces $\sigma=0, L$ and $\sigma=\sigma_0$ is
 given by
\begin{equation}
p^T=p^L-p^E=\rho^L+\rho^E,
\end{equation}
which is generally nonzero according to Eqs.(\ref{Euc}),
(\ref{Lor}) and (\ref{ppp}). This nonzero pressure causes
instability in the location of $\sigma_0$ relative to $\sigma=0$.
Depending on the initial location of $\sigma_0$, the corresponding
value and sign of the pressure may lead one of the regions ( L or
E ) to grow or shrink. It is very hard to judge about the exact
behavior of the pressure from Eqs.(\ref{Euc}), (\ref{Lor}),
because it depends on $N$, the location of $\sigma_0$, and the
complicate functions $(a/b)_\omega$,$(c/b)_\omega$, $(1/b)_\omega$
in which the energy spectrum $\omega$, itself, depends on
$\sigma_0$ through the quantization condition (\ref{11}).

Nevertheless, one may evaluate the situation in the two limits of
$\sigma_0$. In the limit $\sigma_0 \rightarrow 0$, the term
$\frac{\pi}{6(L-\sigma_0)^2}$ may be neglected in comparison with
two other terms. Therefore, there is a competition between the
first sums and third terms in Eqs.(\ref{Euc}), (\ref{Lor}). And,
upon this competition the pressure may cause the Euclidean region
to grow or shrink. On the other hand, in the limit $\sigma_0
\rightarrow L$, the term $\frac{\pi}{6(L-\sigma_0)^2}$ may
dominate the other two terms and the pressure $p^T=\rho^L+\rho^E$
becomes negative, $p^L<p^E$, which means the Euclidean region is
growing ( with increasing pressure ) toward $\sigma_0=L$.
Fortunately, in this case, there is no divergency problem at
$\sigma_0=L$. This is because, once the circle is completely
covered  by Euclidean metric, the quantization condition and all
subsequent calculations break down.

The nonzero pressure obtained above and the consequent change in
the signature changing region will certainly change the energy
spectrum of the scalar fields through the quantization condition
$$
\cosh {\omega \sigma_0}\: \cos {\omega (\sigma_0 - L)} =1.
$$
The modified signature changing region $\sigma_0$ and energy
spectrum $\omega$ back react on the pressure through
Eqs.(\ref{Euc}), (\ref{Lor}). The central term of the algebra
corresponding to infinite conserved charges \cite{far}
$$
[L_\omega,L_{\omega^\prime}] =  (\omega - \omega')
L_{\omega+\omega'} + C(\omega,\omega')
$$
is correspondingly changed through
$$
C(\omega,\omega') = \delta_{{\omega+\omega'},0} f(\omega , \omega'
,\sigma_0) -4\sum_{\omega_1 ,\omega_2 >0 }^N {\omega_1}^2
{\omega_2}^2 C_{\omega_1,\omega_2}^{-\omega'}\: C_{\omega_1
,\omega_2}^{\omega} + \sum_{\omega_1>0}^N {\omega_1}^2\:(\omega -
\omega_1)^2\:C_{\omega_1,\omega-\omega_1}^{-\omega'}
$$
where
$$
f(\omega,\omega',\sigma_0)=3\sum_{l=-n}^0
[-2(l+k)a-\omega-\omega'][2(l+k)a+\omega]^2[|(2(l+k)a+\omega+\omega')(2(l+k)a)|]^{1/2}
$$
and
$$
N=(2k-1)a , \hspace{5mm}  \omega=(2n-1)a, \hspace{5mm}
a=\frac{\pi}{2(2\pi-\sigma_0)}
$$
with $k$ and $n$ as integers.

It is seen that in the special case $N=0$ the first sums and the
last terms vanish in (\ref{Euc}), (\ref{Lor}) and these lead to
the standard result $-\frac{\pi}{6L^2}$ for the pure Lorentzian
metric $\sigma_0=0$ on the cylinder \cite{db}. In fact, $N=0$
corresponds to $\omega=N=0$ which means $\omega$ is an integer
starting from zero; a case which occurs only in the pure
Lorentzian region.

\section{Conclusion}

We have studied a two-dimensional model in which the spacetime is
a cylinder (circle $\times$ real number) with the circle
representing {\it space} and the real line representing {\it
time}. Moreover, we have assumed that this manifold admits a
signature change of the type which had already been reported in
\cite{far}.

We were interested in studying the Casimir effect for the real
massless scalar fields propagating over this manifold. To this
end, we have considered the expressions for the components of
energy-momentum tensors associated with the real scalar field and
calculated the corresponding vacuum expectation values. These
expressions are found to be infinite, hence a regularization
scheme is used to make them finite. By introducing a convenient
cut-off and a regularization scheme, we obtain the finite
expressions for the vacuum expectation values of the energy
momentum tensors. These provide us with the finite energy
densities and pressures in both Euclidean and Lorentzian regions
so that the net pressure on the signature changing hypersurfaces
is obtained. This pressure causes instability in the signature
changing region $\sigma_0$ and this instability alters the energy
spectrum through the quantization condition. The modified
$\sigma_0$ and spectrum $\omega$ themselves back react on the
pressure through Eqs. (\ref{Euc}), (\ref{Lor}). Moreover, the
central term of diffeomorphism algebra of real massless scalar
fields obtained in \cite{far} is altered due to modifications in
$\sigma_0$ and spectrum $\omega$.

The action for free massless scalar field propagating on the
signature changing background
$$
S=\frac{1}{2} \int dt \int_{Lorentzian} d\sigma \sqrt{|g|}
g_L^{\mu \nu} \partial_{\mu} \phi^L_\omega \partial_{\nu}
\phi^L_\omega + \frac{1}{2} \int dt \int_{Euclidean} d\sigma
\sqrt{|g|} g_E^{\mu \nu} \partial_{\mu} \phi^E_\omega
\partial_{\nu} \phi^E_\omega
$$
may be rewritten in the form of string action
$$
S \sim \frac{1}{2} \int dt \int d\sigma \sqrt{|g|} g^{\mu \nu}
\partial_{\mu} \phi^a_\omega \partial_{\nu} \phi^b_\omega \eta_{a
b}
$$
with the distribution $g_{\mu \nu}=\Theta^+ g_{\mu \nu}^E
+\Theta^- g_{\mu \nu}^L$ and
$\phi^a_\omega=(\phi^E_\omega,\phi^L_\omega)$ with $\eta_{a
b}=diag(\Theta^+,\Theta^-)$. In this way it looks like we have a
closed string with Euclidean and Lorentzian parts propagating in a
distributional way in the two-dimensional target space
($\phi^E_\omega , \phi^L_\omega$) \cite{far}. The discontinuous
nature of the model in classifying Euclidean and Lorentzian
solutions $\Phi^E_\omega , \Phi^L_\omega$ with discrete symmetry
under $\omega \leftrightarrow -\omega$ in each class motivates one
to study it in the context of {\it orbifolds}. For example if we
suppose the target space $M$ to be
$\phi^a_\omega=(\phi^E_\omega,\phi^E_{-\omega},\phi^L_\omega,\phi^L_{-\omega})$
and assume a permutation of it $\pi=(\phi^E_\omega \:
\phi^E_{-\omega}) (\phi^L_\omega \: \phi^L_{-\omega})$, then
regarding the definition of an orbifold as the object one obtains
by dividing a manifold by the action of a discrete group, it seems
to be possible to define an orbifold $M/\pi$ which results in
$\phi^a_\omega=(\phi^E_\omega , \phi^L_\omega)$. In this way,
perhaps at a formal level, we may have {\it a string on an
orbifold} \cite{far}.

Therefore, the study of Casimir effect in the present model may
provide important results relevant to the study of closed bosonic
strings. It is also appealing to proceed with the idea of Casimir
effect in different 3+1 dimensional signature changing spacetimes
to investigate what secondary effects may be produced by the
Casimir effect \cite{SD}.

\section*{Acknowledgment}

This work has been financially supported by the Research
Department of Azarbaijan University of Tarbiat Moallem, Tabriz,
Iran.

\end{document}